\documentstyle[aps,multicol,psfig]{revtex}
\begin{document}
\def\pa{\parallel}
\def\pe{\bot}
\def\bea{\begin{eqnarray}}
\def\eea{\end{eqnarray}}
\def\be{\begin{equation}}
\def\ee{\end{equation}}
\let\a=\alpha \let\b=\beta  \let\c=\chi \let\d=\delta  \let\e=\varepsilon
\let\f=\varphi \let\g=\gamma \let\h=\eta \let\k=\kappa  \let\l=\lambda
\let\m=\mu   \let\n=\nu   \let\o=\omega    \let\p=\pi
\let\r=\varrho  \let\s=\sigma \let\t=\tau   \let\th=\vartheta
\let\y=\upsilon \let\x=\xi \let\z=\zeta
\let\D=\Delta \let\F=\Phi  \let\G=\Gamma  \let\L=\Lambda \let\Th=\Theta
\let\O=\Omega
\newcommand{\ie}{\hbox{\it i.e.\ }}         
\draft
\tightenlines
 
\title{On the universality classes of driven lattice gases}
\author{
Pedro L. Garrido $^1$, Miguel A. Mu\~noz$^{1}$ and F. de los Santos $^{1,2}$}
\address{
$^1$ Institute {\em Carlos I} for Theoretical and Computational Physics\\
and Departamento de Electromagnetismo y F\'\i sica de la Materia,
Universidad de Granada,
 18071 Granada, Spain.\\    
$^2$  Universidade de Lisboa, 
 Complexo Interdisciplinar,
 Avenida Profesor Gama Pinto 2,  1649-003 Lisboa, Portugal 
}
\date{\today}

\maketitle
\begin{abstract}
 Motivated by some recent criticisms to our alternative Langevin equation
for driven lattice gases (DLG) under an infinitely large driving 
field, we revisit the derivation of such an equation, and test
its validity. As a result, an additional term, coming from a careful   
 consideration of entropic contributions, is added to the equation.
This term heals all the recently reported generic infrared singularities.
 The emerging equation is then identical to that describing
 randomly driven diffusive systems.
This fact confirms our claim that
 the infinite driving limit is singular,
 and that the main relevant ingredient
determining the critical behavior of the DLG in this limit
is the anisotropy and not the presence of a current. 
Different  aspects
of our picture are discussed, and  it is 
concluded that it constitutes a very plausible scenario to rationalize
the critical behavior of the DLG and variants of it.
\end{abstract}

\pacs{PACS numbers: 64.60.-i, 05.70.Fh}

\begin{multicols}{2}
\narrowtext

The  driven lattice gas (DLG) \cite{katz,zia,marro}
is a simple nontrivial extension of the kinetic Ising model,
and constitutes certainly a main archetype of out-of-equilibrium system. 
Fully understanding the critical properties of the DLG
would be a fundamental milestone on the way to rationalizing
the fast developing field of nonequilibrium phase transitions. 
The DLG is defined as a half filled, d-dimensional kinetic Ising model 
with conserved dynamics, in which transitions  
in the direction (against the direction) of an external 
 field, $E$, are favored (unfavored) \cite{katz,zia,marro}.
The external field induces two main nonequilibrium effects:  
the presence of a 
net current of particles in its direction, and 
 anisotropic system configurations.
 At high temperatures the system is in a disordered phase, while   
below a certain critical point it
orders by segregating into high and low density 
aligned-with-the-field stripes.

In order to analyze the DLG critical nature, and determine its degree
of universality, a Langevin equation intended to capture
the relevant physics at criticality was proposed and renormalized
more than a decade ago \cite{JS}.
 This elegant theory, the {\it 
driven diffusive system} (DDS) seems to capture
the main symmetries and conservation laws of the discrete DLG
(including a current term as the most relevant nonlinearity),
and is therefore a suitable and very reasonable
candidate to be {\it the} 
canonical continuous model, representative of the DLG universality
class.
 
Unfortunately, the most emblematic prediction coming from the
analysis of the DDS equation, namely, the mean field behavior
of the order parameter critical exponent ($\beta=1/2$ \cite{JS}),
has not been compellingly verified in any Monte Carlo simulation 
of the DLG
in spite of the huge computational effort devoted to test it.
 In particular, systematic deviations from scaling are observed   
both in $d=2$ \cite{marro,FSS} and in $d=3$ \cite{LW}
if data collapse is attempted using $\beta=1/2$ \cite{aclarar}.
On the other hand, different Monte Carlo numerical simulations 
(performed in different geometries
and using different finite size scaling ansatzs) 
lead systematically to a value of $\beta$ around $0.3$
with error bars apparently excluding $\beta=1/2$
(we refer the interested reader to 
\cite{marro} for a review of simulation analysis).
This is a main indication that the DDS Langevin equation
 does not describe properly the DLG at criticality.

Moreover, there are some other hints suggesting strongly that
the discrepancies between the predictions of the standard theory
and Monte Carlo results are more fundamental than
a simple numerical difference in $\beta$.
In particular, the intuition developed from 
Monte Carlo simulations of the DLG and variants of it \cite{marro,RDDS2}
(performed under large external driving fields)
suggests 
that, contrarily to what happens for the DDS equation,
 it is the anisotropy and not the presence
of a current the most relevant ingredient for criticality.
For instance, in a modification of 
the DLG in which anisotropy is included   
by means others than a current \cite{ALGA},
the scaling behavior at 
criticality remains unaltered upon the switching on of
an infinite driving 
(see the appendix and \cite{ALGA,marro}).
 Other compelling evidences supporting this hypothesis can be found
in \cite{marro,MAGA}.

In order to shed some light on this puzzling situation 
and reconcile theory with numerics,
different possible scenarios have been explored; but
so far no satisfactory clarification has been reached.
Within this context, we have recently 
revisited the time-honored DDS equation and questioned
its general validity \cite{PPY,PP}. 
 In particular, we have tackled the task of constructing 
a coarse-grained procedure in a more detailed way such that, starting 
from a Master equation representing the  DLG, would              
give as output a continuous Langevin equation. This approach 
permits us to keep track of microscopic details that 
could eventually be overlooked when writing down a 
Langevin equation 
respecting naively the microscopic symmetries and conservation constraints
\cite{JS}.  
 This approach has given rise to a rather unexpected and quite 
interesting output: The limit of infinitely large driving 
(i.e. the limit in which attempted jumps in the direction 
of the field
are performed with probability one and jumps against E are strictly forbidden)
is singular \cite{PPY}.  Let us stress that in order 
to enhance nonequilibrium effects
most of the available computer studies are performed in this limit.
The main results derived so far using our approach are:  
\begin{itemize}
\item For vanishing values of the driving field it leads to the
standard equilibrium model B \cite{HH}, capturing the relevant physics
of the kinetic Ising model with conserved dynamics. 
\item For nonvanishing, but {\it finite } driving fields we reproduce 
the standard DDS Langevin equation \cite{zia,JS}.
\item
 In the  limit of infinitely large driving,   
 where the dependence of jumps in the direction of the field 
on energetics is replaced by a zero-one (all-or-nothing) condition,
a different Langevin equation emerges.
This new equation has the main property of not including 
any relevant term coupling $E$ to the density field $\phi$ \cite{PPY},
and the presence of anisotropy is its main relevant ingredient.
\end{itemize}

 The new Langevin equation for the DLG under infinitely large driving
proposed in \cite{PPY,PP} and renormalized
in \cite{PY} has some important virtues  to be discussed afterwards,     
but seems also to exhibit some pathologies, as 
 recently 
pointed out by Caracciolo et al. \cite{Pisa} and  
also by Schmittmann et al. \cite{micidiale}.
In what follows we show that such anomalies can be 
healed in a rather natural way,
and do not disprove at all the general validity of  our
new approach (as could be inferred from \cite{micidiale}). 

 Let us now present the Langevin equation derived in \cite{PPY,PP}    
for the infinite driving limit, report on its deficiencies
\cite{Pisa,micidiale}, and discuss the way to heal them.
 The equation reads \cite{PPY,PP}:  
\bea
\label{newold}
\partial_t \phi ={e_0 \over 2}
\Big[ -\D_\pa \D_\pe \phi-\D_\pe^2 \phi+
\t \D_\pe \phi+ {g \over 3!} \D_\pe \phi^3 \Big] \nonumber \\
+\sqrt{e_0} \ \nabla_\pe \cdot \mbox{\boldmath $\xi$}_\pe
+\sqrt{e_0 \over 2} \ \nabla_\pa \xi_\pa,
\eea
where $\nabla_\pa $ ($\nabla_\pe$) is the gradient operator in    
the direction parallel (perpendicular) to the driving field, and
 $\xi$ is a conserved Gaussian white noise \cite{PPY,PY}.
This equation is analogous to a model B in the direction(s) perpendicular
to the field,
coupled to a simple random diffusion mechanism in the
parallel direction. The origin of all the difficulties pointed out 
in \cite{Pisa,micidiale} can be  traced back to the following
property:
Defining the total density for each value of $r_\pa$, 
$\Upsilon (r_\pa,t) \equiv \int d^{d-1} r_\pe \phi(r_\pa, {\bf r_\pe})$,
it is not difficult to see (after averaging over the noise) that
$\Upsilon (r_\pa)$ 
is a conserved quantity for all values of $r_\pa$ \cite{Pisa}.
 Observe also that $\Upsilon (r_\pa)$ is
nothing but the zero Fourier mode of the density at each column.
These (spurious) conservation laws, absent in the DLG, are at the origin of
the infrared singularities appearing in 
Eq. (\ref{newold}) \cite{Pisa,micidiale}.
 
In order to investigate the causes of this deficiency in our Langevin
equation
and eventually  overcome the problem of the extra conservation laws and
associated infrared divergences,
 we have re-analyzed our derivation of Eq. (\ref{newold})
in \cite{PPY,PP}. One can easily see that the transition
rates in the microscopic master equation in \cite{PPY,PP} were written as
depending on the variations of two adding contributions:
the free energy functional (the usual Ginzsburg-Landau free energy)
and the  external driving-field contribution.
  The transition rates, written in that way,
saturate to zero or one in the field direction,  
 in the limit of infinite driving.
This saturation erases any further dependence on the free energy density
(which includes both entropic and energetic contributions).
On the contrary, in the DLG it is only the dependence
on the Ising energetics that becomes negligible
in the limit of large driving fields. 
In a coarse grained description we should therefore separate energetic 
from entropic terms. With this guiding idea, 
we have reconsidered our derivation of Eq. (\ref{newold}) 
 and rewritten
the transition
rates in \cite{PPY,PP} as the product of two 
contributions: one controlling the energetics and the other one the entropic 
part \cite{parisi}. 
By performing a calculation analogous to that in \cite{PPY,PP},
but including the transition rates written in this modified way, 
it is a matter of algebra to see that
 a new term (missing in \cite{PPY,PP,PY}) emerges:
$\rho \nabla_\pa \phi({\bf x},t)$  \cite{details}.

It is straightforward to verify that apart from properly keeping track
of entropic contributions, this extra (mass) term 
 heals all of the aforementioned problems in Eq. (\ref{newold}): 
 no spurious conservation laws are involved and
generic infrared singularities disappear.
 
 Let us now discuss how this new additional
 term affects the results presented in \cite{PY}.
 Performing a naive scaling analysis, one sees
that $x_\pa \sim {x_\pe}^2$ \cite{anisotropy}, and upon elimination 
of naively irrelevant terms and absorbing $e_0$ into the time scale,
 one obtains our final result:
the critical Langevin theory under infinitely large driving
\be
\partial_t \phi =
   \rho \D_\pa \phi  -\D_\pe^2 \phi+
\t \D_\pe \phi+ {g \over 3!} \D_\pe \phi^3 
+ {2 \over \sqrt{e_0}} \nabla_\pe \cdot \mbox{\boldmath $\xi$}_\pe,
\label{ADS}
\ee                                               
that we call the
{\it anisotropic diffusive system} (ADS).
 This turns out to be a well known Langevin equation:  
 the continuous 
representation of the randomly driven DLG \cite{RDDS,zia},
 i.e. a  DLG in which
the external field changes sign randomly in an unbiased fashion.
 The main difference 
between this theory and the DDS is that the ADS does not include
an overall current.
The current term $E \nabla_\pa \phi^2$ appearing in the DDS (and constituting
 its most relevant nonlinearity)
is absent here. In the random DLG, 
such a term cannot appear for symmetry reasons, while in the infinite 
driving case discussed in this paper 
it is the saturation of the transition rates in the field direction
that prevents such a current term from appearing. 
%For finite driving it 
%re-emerges as the leading nonlinearity.

 The cubic operator and the Laplacian term in the parallel direction
in Eq. (\ref{ADS})
are both marginal at the critical dimension $d=3$. The
results up to first order in an epsilon expansion
of Eq. (\ref{ADS}) around $d=3$  are \cite{RDDS,same}:    
$\nu_\pe = {1/2 + \e/12}$, and 
$\beta=1/2-\epsilon/6$ \cite{same}.
 Observe that in $d=2$ one obtains $\beta=1/3$ (slightly modified 
by two-loop corrections \cite{RDDS})
in remarkable good agreement with  Monte Carlo results.
 For instance (see table and \cite{nu}):
 the best available Monte Carlo result
for the random DLG is $\beta \approx 0.33$ \cite{RDDS}; for the infinitely 
driven DLG $\beta \approx 0.30 \pm 0.05$  \cite{marro};
 and $\beta \approx 0.34$ for the closely related model studied in
\cite{ALGA}, called ALGA, and argued to belong to the same universality class 
(see appendix).

 Some further comments on the validity of our approach follow:
(i) Our complete theory (including constant and irrelevant terms) does
have a net current \cite{PPY,PP}, though it does not
enter the final Langevin equation.
(ii) An infinitely large field is, in practice, any for which transitions
against the field never occur. Given that all
commonly used transition rates depend on $E$ through exponential
functions, field
values much larger that unity can be considered infinite for all
practical purposes in Monte Carlo experiments.
  For smaller fields,
 we expect crossover effects from the infinite field regime
(ruled by the ADS)
to the finite-driving standard DDS behavior to occur.
 These crossovers could
obscure the numerical observation of the DDS mean-field exponent beta for
large but finite driving fields.               
(iii) The introduction of the new term in the parallel direction 
heals all the possible problems in relation to infrared singularities,
extra conservation laws, and
 anomalies in the structure function \cite{micidiale}.
In particular, the structure function presents a discontinuity 
singularity as happens in the DLG  \cite{zia}.
(iv) Given the absence of any relevant current term in  Eq. (\ref{ADS}),
 the critical theory has ``up-down''
symmetry ($\phi \rightarrow -\phi$). This symmetry, in principle,
 is absent in 
the microscopic model as the presence of nonvanishing
three-point correlation functions seems to indicate \cite{zia}. 
However, it is not clear whether  
such correlations are relevant at criticality or not. 
As an indication that in fact they could well be irrelevant,
we discuss here 
the problem of triangular anisotropies \cite{triangles}: 
In both the DLG and the DDS, 
 droplets 
of the minority phase (if any) 
develop triangular shapes, 
closely related to the existence of nonvanishing three-point
 correlations.
 However,
triangles orientate in opposite directions in the 
microscopic DLG and in the continuous DDS. This difference
seems to be not universal as shown  by recent Monte Carlo studies  
\cite{triangles},
i.e. it depends on microscopic details and can be modified by changing them
both in the DLG and in the DDS.
This fact supports the idea that 
nonvanishing three-point correlation functions
might not be a relevant ingredient for a description of the DLG at criticality.
More significatively, {\it simulations show that
for large enough driving fields the
triangular anisotropy is suppressed} (see \cite{triangles}), 
 providing an indication that the up-down symmetry is
restored in the infinite driving limit. This constitutes, 
we believe, another strong backing of our picture.  
 
  In summary,  we have discussed the plausibility of the alternative 
field theoretical approaches to driven lattice gases under the
effect of an infinitely large external driving field. Some deficiencies
recently pointed out are overcome by 
introducing an extra Laplacian term in the direction of the field
 in the Langevin equation first proposed in \cite{PPY,PP}. 
 This new term, 
coming from a proper consideration of entropic contributions,
had been overlooked in previous papers.
Our approach
leads to the following global picture:
(i) For $E=0$, model B reproduces the equilibrium critical properties of
isotropic diffusive systems.
(ii) For finite driving field, the standard DDS Langevin equation, including
a current term, should
describe properly the long wavelength properties around the critical point.
(iii) The limit of infinitely fast driving is singular: {\it the current
is irrelevant, and the anisotropy becomes the main relevant property.}
The leading critical properties in this case are expected to be described
by the ADS, Eq. (\ref{ADS}).
 The reason for this being that in the presence of
infinite driving the transition rates saturate to 1 (0) for allowed
 (forbidden) transitions in the driving direction, and no further
track of coupling between E and the density field survives in the 
resulting Langevin equation. We want to stress at this point that 
this property was not obvious a priori, but emerges as a natural
 output from our model building strategy.
 
The proposed Langevin equation for large external field 
 provides a quite plausible scenario
shedding some light on a difficult problem. In particular, it
justifies the observed lack of differences (for large fields)
 in simulations in systems with and
without a current, and provides a likely 
justification of why the standard prediction $\beta=1/2$ is not
confirmed in Monte Carlo simulations, and instead a value 
$\beta \approx 0.33$ is observed.

  In order to test numerically the picture presented in this
paper, it would be highly desirable to perform extensive
simulations for finite driving field ($E \approx 1$), and 
study whether differences with respect to the available Monte Carlo
results for large fields appear.
It would also be interesting to    
improve the finite size scaling analysis following the 
strategy used in \cite{CS,Pisa}.
\vspace{0.2cm}

{\bf \centerline{Appendix.}}
  As an evidence aimed to 
 transmitting the intuition that,  
for infinitely large driving,
the current is not relevant at criticality, let us briefly discuss
in this appendix a rather compelling Monte Carlo observation.
 It corresponds to a
 variation of the DLG, named ALGA (anisotropic
lattice gas automaton); see \cite{ALGA} 
for a detailed definition. This model is placed by definition 
 at the limit of infinite driving:
 jumps in the anisotropy direction are performed
randomly without attending to energetic considerations. Simulations
are performed both in the presence of an overall current
 (case $p\neq 1/2$ in \cite{ALGA})
and in the absence of it ($p=1/2$); 
the curves for the order parameter versus
the distance to the critical point are indistinguishable 
in the cases with and without a current
(figure 3 in \cite{ALGA} is particularly illuminating). 
It could be argued that the details of this modified model  \cite{ALGA}
 render it not completely equivalent to the original DLG.
 However, we do not think these microscopic differences have any relevance
at a coarse grained level. 
In fact, 
we expect this model to be represented
by Eq. (\ref{ADS}): in one direction
particles tend to stay together, and it is natural to 
assume that their coarse grained behavior is controlled
 by a model B in this
direction. In the other direction, jumps occur regardless of energetics
and, therefore, the dynamics becomes purely diffusive. With these two
ingredients we recover the ADS, Eq. (\ref{ADS}),
 as the Langevin equation for the ALGA.
As a further evidence  supporting this hypothesis  let us mention
that the measured $\beta$ exponent in the ALGA is $\beta \approx 0.34$
 (again very close to the value $\beta \approx 1/3$)  
in both cases: with and without current.

{\it ACKNOWLEDGMENTS-}
 It is a pleasure to acknowledge J. Marro
and J. L. Lebowitz 
for useful discussions and encouragement.
 We thank with special gratitude      
S. Caracciolo and collaborators for sharing with us extremely useful
and valuable unpublished results.
 This
work has been partially supported by the
European Network Contract ERBFMRXCT980183
 and by the Ministerio de Educaci\'on
under project DGESEIC, PB97-0842.

\begin{table}
\begin{center}
\begin{tabular}{|c|c|c|c|c|}
\hline
$ MODEL$  && No net current &  Net current   & Net current  \\
          &&    & Finite driving & Infinite driv. \\
\hline
\hline
 DLG && Model B  & DDS    & ADS  \\
     &&  $\beta = 0.125$  & & $ \beta \approx 0.3$ \\
 \hline
 Random DLG  && ADS  &  DDS   & ADS  \\
     && $ \beta \approx 0.33$  & &  \\
\hline
 ALGA  && ADS  & Undefined & ADS \\
     && $ \beta \approx 0.34 $ &        & $ \beta \approx 0.34$ \\
\hline
\end{tabular}
\end{center}
\label{tabla}
\noindent{\small Universality classes for different models as
a function of the driving field value.  The  reported values of $\beta$
correspond to the best-to-date numerical results in $d=2$.
 The theoretically
predicted value is $\beta=1/2$ for the DDS and 
$\beta \approx 0.33$ for the ADS.}  
\end{table}
                                           
\newpage
\end{multicols}

\begin{thebibliography}{99}
\bibitem{aclarar} Let us stress that other predictions of the standard
field theory seem to be verified by performing a finite size
scaling analysis (see \cite{FSS} and \cite{LW}), but a fully consistent
scaling, including good data collapse for the order parameter,
has never been observed. 
See the criticisms to \cite{FSS} and \cite{LW}
in \cite{MAGA}.
\bibitem{details} The detailed calculation will be presented elsewhere.
\bibitem{anisotropy}  Observe that in the  absence of the new term
isotropic naive scaling, $x_\pa \sim x_\pe$ had to be assumed.
 In the present case, $x_\pa \sim x_\pe^2$ is obtained
by imposing the leading term in the parallel direction to be marginal
at the critical dimension.
\bibitem{same} Note that results up to first order in epsilon
coincide with
those obtained in our previous paper \cite{PY}, but here $\e=3-d$.
\bibitem{nu} The prediction for the exponent $\nu$ up to two loops
is $\nu\approx 0.63$. Monte Carlo simulations
 yield values
around $0.7$ for the DLG \cite{marro}, and $\nu \approx 0.6$
for the random DLG \cite{zia}.
\end{thebibliography}
\end{document}